\documentstyle[11pt,aaspp4]{article}

\slugcomment{Submitted to the Astrophys. J.}

\lefthead{Levshakov, Takahara \& Agafonova}
\righthead{Kinetic temperature measurements}

\begin{document}

\title{ Measurability of kinetic temperature from metal
absorption-line spectra formed in chaotic media\altaffilmark{1}}

\author{Sergei A. Levshakov}
\affil{National Astronomical Observatory, Mitaka, Tokyo 181, Japan}

\author{Fumio Takahara}
\affil{Department of Earth and Space Science, Faculty of Science,
Osaka University, Toyonaka, Osaka 560, Japan}

\and

\author{Irina I. Agafonova}
\affil{National Astronomical Observatory, Mitaka, Tokyo 181, Japan}

\altaffiltext{1}{Based in part on data obtained at the W. M. Keck Observatory,
which is jointly operated by the University of California and the California Institute
of Technology.}

\begin{abstract}
We present a new method for recovering the kinetic temperature
of the intervening diffuse gas to an accuracy of 10\%.
The method is based on the comparison of unsaturated absorption-line profiles
of two species with different atomic weights.
The species are assumed to have the same
temperature and bulk motion within the absorbing region.
The computational technique involves the Fourier transform of the
absorption profiles and the consequent Entropy-Regularized $\chi^2$-Minimization [ERM]
to estimate the model parameters. The procedure is tested using synthetic spectra
of C$^+$, Si$^+$ and Fe$^+$ ions.
The comparison with the standard Voigt fitting
analysis is performed and it is shown that the Voigt deconvolution of the
complex absorption-line profiles may result in estimated temperatures which are not physical.
We also successfully analyze Keck telescope spectra of
C~II$\lambda1334$ and Si~II$\lambda1260$ lines observed 
at the redshift $z = 3.572$ toward 
the quasar Q1937--1009 by Tytler {\it et al.}. 
\end{abstract}


\section{Introduction}

New generation of large telescopes enables to study the intergalactic 
and interstellar absorbing gas 
with high instrumental resolutions sufficient to resolve
absorption features into the individual subcomponents. 
From absorption-line spectra, one usually estimates the column densities of
absorbing species, $N$, the Doppler $b$-parameter of the diffuse gas, 
and the positions (i.e. radial velocities) of the subcomponents 
which are traditionally considered as being caused by the intervening `clouds'.

The observations show the ubiquitous asymmetry of the line profiles  
(Spitzer \& Fitzpatrick 1993; Huang {\it et al.} 1995;
Lu {\it et al.} 1996; Prochaska \& Wolfe 1997).
Their widths are believed to reflect the sum of thermal and non thermal
(`bulk' or `turbulent') motion. It should be noted
that the term `turbulent' used in spectroscopy is less distinct as
compared with hydrodynamic turbulence and labels the {\it lack} of information
about the nature of the broadening mechanism (Traving 1975). Therefore
to treat observations a model of the line broadening is to be
specified first of all.

It is a commonly used assumption that both thermal and turbulent velocities
along a given line of sight can be represented by Gaussian distributions, 
i.e. thermal and non thermal broadening can be added quadratically 
\begin{equation}
b^2 = v^2_{\rm th} + 2\sigma^2_{\rm t}\ .
\label{eq:E1}
\end{equation}
Here $\sigma_t$ is the rms turbulent velocity and $v_{\rm th}$ is the thermal velocity  
\begin{equation}
v_{\rm th} = \sqrt{2k_{\rm B}T/m_{\rm a}}\: ,
\label{eq:E2}
\end{equation}
where $k_{\rm B}$ is Boltzman's constant, $m_{\rm a}$ is the mass of the element, 
and $T$ the kinetic temperature.

To separate the contribution of thermal and turbulent motion one usually utilizes the 
line profiles of different elements assuming that they have the same temperature and the 
velocity distribution within the individual components.
If this is the case, then higher mass 
elements should have lower apparent $b$-values 
provided the measurements have a sufficiently high spectral resolution.
With such reservations one can 
solve equation~(\ref{eq:E1}) for $T$ 
and $\sigma_t$  for different species.

Applied to the real absorption-line spectra, this technique often leads, however, 
to the implausible wide range of $T$ among individual `cloudlets' 
yielding for some of them even unphysical negative kinetic temperatures --
clearly an absurdity 
(Spitzer \& Fitzpatrick 1993, 1995; 
Fitzpatrick \& Spitzer 1997;  Tripp, Lu \& Savage 1997). 
This suggests that in general the problem of the $b$-parameter measurement 
from the interstellar 
and/or quasar metal absorption-line spectra is {\it far from being solved}.

To a certain extent this problem is connected with the interpretation
of complex absorption features as being caused mainly by the
density fluctuations.
However, as shown by Levshakov \& Kegel (1997)
the intensity fluctuations within the line profile can occur due to
correlation effects in the large scale velocity field even for the
homogeneous gas density.

The present paper is primarily aimed at the problem of the kinetic 
temperature measurement accounting for these correlation effects.
The analysis employs absorption profiles of two species with different
atomic weights. 
These profiles are considered to arise in 
a transparent absorbing region of thickness $L$ 
with a correlated velocity field and a homogeneous temperature $T$.
Although in principle density fluctuations can be incorporated,
here we confine ourselves to the case of the homogeneous gas density
and, hence, the density ratio of the chosen elements is constant along
the line of sight.
These conditions yield the similarity in the observed profiles
of different elements
(examples can be found in Spitzer \& Fitzpatrick 1993, or in Songaila 1998).
The assumption of constant $T$ is supported by the fact that 
the states of thermally stable, optically thin interstellar gas 
take only limited temperature ranges over a relatively wide range of
density variations (see, e.g., Figure~8.9 in Shu 1992). 

The Entropy-Regularized $\chi^2$-Minimization (the ERM method) 
we propose allows a robust measurement of a single value of $T$ from
the multicomponent (complex) absorption spectra 
in cases where the standard Voigt-profile fitting procedure yields
different values of $T$ for the different subcomponents. The description
of our model and its basic assumptions is given in \S~2. 
A study of the accuracy of the method
is provided through a series of numerical simulations of complex spectra
produced in the correlated velocity fields of different structure which
is presented in \S~3. We proceed further in this approach with high
resolution  C~II$\lambda1334$ and Fe~II$\lambda2600$
synthetic spectra in \S~4, in order to derive the gas temperature and to compare its value
with the standard Voigt fitting procedure yield. 
An example based on real observations of high-redshifted metal absorption lines 
is presented in this section as well.  We summarize our conclusions in \S~5.

\section{Data analysis technique}

In this section we present a compendium of the standard radiative transfer
equations describing pure absorption in spectral lines, and the consecutive
analytical technique to derive the $T$-estimation and the radial-velocity
distribution function $\phi_{\rm t}$ from the line profiles of two
elements with different atomic weights. Here, we are dealing with noiseless
data, assuming that the given spectra are exact.

\subsection{Basic equations}

If the interstellar (intergalactic) absorption line is optically thin,
one observes directly the apparent optical depth 
$\tau^\ast(\lambda)$ as function of wavelength $\lambda$ 
\begin{equation}
\tau^\ast(\lambda) = \ln \left[ I_0(\lambda)/I_{\rm obs}(\lambda) \right]\: ,
\label{eq:E4}
\end{equation}
where $I_{\rm obs}(\lambda)$ and $I_0(\lambda)$ are the intensities with and
without the absorption, respectively.

The recorded spectrum is a convolution of the true spectrum and the
spectrograph point-spread function $\phi_{\rm sp}$
\begin{equation}
I_{\rm obs}(\lambda) = \int^{+\infty}_{-\infty}\,I_0(\lambda')\,
{\rm e}^{-\tau(\lambda')}\,\phi_{\rm sp}(\lambda - \lambda')\,{\rm d}\lambda'\: ,
\label{eq:E5}
\end{equation}
where $\tau(\lambda)$ is the true (intrinsic) optical depth.

The intensity of the background continuum source $I_0(\lambda)$
changes usually very slowly over the width of the spectrograph function,
and hence~(\ref{eq:E5}) becomes
\begin{equation}
I_{\rm obs}(\lambda) = I_0(\lambda)\,\int^{+\infty}_{-\infty}\,
{\rm e}^{-\tau(\lambda')}\,\phi_{\rm sp}(\lambda - \lambda')\,{\rm d}\lambda'\: .
\label{eq:E6}
\end{equation}

For the following it is convenient to change from $\lambda$-space to the 
radial-velocity $v$-space.  
Further, let the line-absorption coefficient reckoned per atom
(the absorption cross-section) at radial velocity $v$ 
within the line profile be denoted by $k(v)$.
Then the optical depth at velocity $v$ is
\begin{equation}
\tau(v) = N\,k(v)\: .
\label{eq:E7}
\end{equation}

For the case of unsaturated absorption lines when the effect of 
radiation damping is negligible $k(v)$ may
be described by the Doppler kernel of the Voigt function (Spitzer 1978).
Then the profile function is defined {\it locally} by the thermal broadening, i.e.
the linewidth is determined by the kinetic temperature only ($b = v_{\rm th}$). 
Now $k(v)$ can be expressed as the product of a constant $k_0$ for a particular 
line and the Gaussian line profile function $\phi_{\rm th}(v)$ 
\begin{equation}
k(v) \equiv k_0 \phi_{\rm th}(v) = \frac{\pi e^2}{m_e c}f_{\rm abs}\lambda_0\,\frac{1}
{\sqrt{\pi} v_{\rm th}}{\rm e}^{ -(v-v_0)^2/v^2_{\rm th}}\: ,
\label{eq:E8}
\end{equation}
where $m_e$ and $e$ are the mass and charge of the electron,
$f_{\rm abs}$ and $\lambda_0$ are the oscillator strength and the wavelength
of the line center at which the corresponding radial velocity is equal to $v_0$.

If atoms are involved in both thermal and hydrodynamic motions 
and the temperature is constant along the line of sight,  
then the total
line profile function is a convolution of $\phi_{\rm th}(v)$ and
the radial-velocity distribution function $\phi_{\rm t}(v)$ 
caused by bulk motion
\begin{equation}
\phi_{\rm tot}(v) = 
\int^{+\infty}_{-\infty}\, \phi_{\rm t}(v') \phi_{\rm th}(v - v')\,{\rm d}v'\: , 
\label{eq:E9}
\end{equation}
and $\tau(v) = N\,k_0\,\phi_{\rm tot}(v)$.

Both functions $\phi_{\rm t}$ and 
$\phi_{\rm th}$ are normalized so that $\int (\cdot)\,{\rm d}v = 1$. 
In principle, $\phi_{\rm t}$ reflects the density fluctuations as well,
and in reality is the density weighted radial-velocity distribution. But while we
are dealing with the similar profiles the explicit treatment of the density 
fluctuations can be avoided.

The instrumental point-spread function $\phi_{\rm sp}$ can be determined
experimentally, whereas $\phi_{\rm t}$ and $\phi_{\rm th}$ 
are the two contributions of astrophysical interest. 

\subsection{$T$ and $\phi_{\rm t}(v)$ calculations}

We now consider how to measure the kinetic temperature $T$
from the complex profiles of a pair of elements with different masses.
It is a task to deconvolve the thermal and the turbulent profiles
both of which are unknown a priori. We are mainly interested in cases
when line profiles are dominated by non thermal broadening, 
$\sigma_{\rm t}/v_{\rm th} \gtrsim 1$, otherwise the problem is trivial. 

Let us assume that two ions have different masses, $m_\iota$, and $m_2 > m_1$.
For a fixed value of $v$ within the line profile the optical depth is 
given according to~~(\ref{eq:E7}-\ref{eq:E9}) by
\begin{equation}
\tau_\iota(v) = \omega_\iota \int^{+\infty}_{-\infty}\,\phi^\iota_{\rm t}(v')\,
\phi^\iota_{\rm th}(v-v')\,{\rm d}v'\; ,\; \iota = 1,2\: ,
\label{eq:E10}
\end{equation}
where $\omega_\iota = k^\iota_0 N_\iota$.

This equation is a convolution of two functions.
Its Fourier transform is 
\begin{equation}
\hat{\tau}_\iota(q) = 
\int^{+\infty}_{-\infty} \tau_\iota(v)\,{\rm e}^{2\pi iqv}\,{\rm d}v =
\omega_\iota\,\hat{\phi}^\iota_{\rm th}(q)\,\hat{\phi}^\iota_{\rm t}(q)\; ,\; 
\iota = 1,2\: ,
\label{eq:E11}
\end{equation}
where $\hat{\phi}^\iota_{\rm th}(q)$ and 
$\hat{\phi}^\iota_{\rm t}(q)$ are the Fourier transforms of the line
profile function $\phi^\iota_{\rm th}(v)$ and 
the radial-velocity distribution function
$\phi^\iota_{\rm t}(v)$, respectively.

If the chosen ions trace the {\it same} volume elements 
and the ionization fraction is {\it equal} for both of them 
(this assumption may not be universally true for any pair of ions but
is expected to be valid for ions with similar ionization potentials),
then $\phi^1_{\rm t}(v) = \phi^2_{\rm t}(v)$. 
In this case two Fourier transforms  $\hat{\tau}_1(q)$  
and $\hat{\tau}_2(q)$ yield the {\it exact} equations relating the line shapes
in the Fourier space to the kinetic temperature {\it independently} of
the velocity and density fluctuations along a given line of sight~:
\begin{eqnarray}
\Re \left[ \frac{\hat{\tau}_2(q)}{\hat{\tau}_1(q)} \right] & = & 
\frac{\omega_2}{\omega_1} {\rm e}^{q^2 \pi^2 (v^2_{\rm th,1}-v^2_{\rm th,2})} \equiv
\frac{\omega_2}{\omega_1} {\rm e}^{a T q^2 \pi^2}\: , \nonumber \\
\Im \left[ \frac{\hat{\tau}_2(q)}{\hat{\tau}_1(q)}\right] & = & 0\: , 
\label{eq:E12}
\end{eqnarray}
where 
$a = 2 k_{\rm B} (m_2 - m_1)/(m_1 m_2)$ 
is a numerical constant.

In the calculation of $T$ from equation~(\ref{eq:E12}) 
the first step is to determine the Fourier transforms 
$\hat{\tau}_{1}$ and $\hat{\tau}_{2}$. 
The most appropriate way would be to represent $\tau$ by a set of
functions which have analytical Fourier transform. We have chosen
Gaussian functions since 
irregular Doppler shifts of the local absorption coefficient
(\ref{eq:E8}) along the sightline, yielding the apparent complexity
of the line profiles, result in a mixture of Gaussians (\ref{eq:E9}). 
The line profiles are modeled using the following sum of Gaussians
\begin{equation}
\tau_\iota(v) = \frac{1}{\sqrt{\pi} \beta_\iota} \sum^{M}_{j=1} 
a_{\iota,j} {\rm e}^{-(v-v_j)^2/\beta^2_\iota}\; ,\; \iota = 1,2\: .
\label{eq:E13}
\end{equation}
Here $M$ is the number of modes and $a_{1,j}$ and $a_{2,j}$ are their weights. 
Since basic assumptions of our method lead to the similarity of both profiles,
we may set the number of modes and their centers to be identical, 
i.e. $M_1 = M_2 = M$ and $v_{1,j} = v_{2,j} = v_j$.
All modes describing the individual profile have the same dispersion, denoted
$\beta_1$ for the first profile and $\beta_2$ for the second,
with $\beta_2 < \beta_1$. 
The choice of the equidispersion Gaussians is by no way restrictive
in our case because the modes in~(\ref{eq:E13}) are used
only to describe the entire shape of each line profile
and do not bear any physical sense as it is in the Voigt profile
fitting procedures. The number of modes $M$ is an arbitrary 
parameter and is determined by
a satisfactory fit of $\tau_1$ and $\tau_2$ solely.  
The equidispersion modes with identical positions allow very easy to meet
the second condition in~(\ref{eq:E12}), -- zero imaginary part of the 
ratio $\hat{\tau}_2/\hat{\tau}_1$, -- resulting in the requirement
$a_{1,j}/a_{2,j} = \xi = {\rm constant}$ for all $j$.

It should be noted that the multi-Gaussian model~(\ref{eq:E13})
is not the unique one to describe the shape of profiles. Any other
representations can be used as well. Nevertheless the model~(\ref{eq:E13})
has been favored because of its clearness and easiness in realization.
The similar profiles can be always fitted by using this model.
And on the contrary, if the fitting according to~(\ref{eq:E13}) fails
then the profiles are non-similar and hence our method cannot be
applied.

Equation~(\ref{eq:E13}) shows that the total number of free parameters to be
found is equal to $2M + 3$, and that $\omega$ can be calculated as
\begin{equation}
\omega_\iota = \sum^{M}_{j=1} a_{\iota,j}\; ,\; \iota = 1,2\: ,
\label{eq:E14}
\end{equation}
with $\omega_2/\omega_1 = 1/\xi$.

Using expansion (\ref{eq:E13}), the Fourier transforms of the line
profiles can be expressed as
\begin{equation}
\hat{\tau}_\iota(q) = {\rm e}^{- \pi^2 \beta^2_\iota q^2}\,\sum^{M}_{j=1}
a_{\iota,j}\,{\rm e}^{2\pi iv_jq}\; ,\; \iota = 1,2\: .
\label{eq:E15}
\end{equation}

From equations~(\ref{eq:E12}--\ref{eq:E15}) it follows that the kinetic 
temperature is determined by
\begin{equation}
T = (\beta^2_1 - \beta^2_2)/a\: .
\label{eq:E16}
\end{equation}

Having estimated the kinetic temperature, the radial-velocity
distribution function $\phi_{\rm t}(v)$ 
can be calculated through the inverse Fourier transform 
\begin{equation}
\phi_{\rm t}(v) = \int^{\infty}_{-\infty} 
\hat{\phi_{\rm t}}(q)\,{\rm e}^{-2\pi ivq}\,{\rm d}q
\label{eq:E17}
\end{equation}
of the function
\begin{equation}
\hat{\phi_{\rm t}}(q) = \frac{\hat{\tau}_\iota(q)}{\omega_\iota\,
\hat{\phi}^\iota_{\rm th}(q)}\; ,\; \iota = 1,2\: .
\label{eq:E18}
\end{equation}
From equations~(\ref{eq:E15}), (\ref{eq:E17}) and (\ref{eq:E18}) we can readily show that
\begin{equation}
\phi_{\rm t}(v) = \frac{1}{\omega_\iota\,\sqrt{\pi (\beta^2_\iota - v^2_{\rm th,\iota})}}
\sum^{M}_{j=1} a_{\iota,j}\,
\exp\left[ - \frac{(v-v_j)^2}{\beta^2_\iota-v^2_{\rm th,\iota}} \right]\; ,
\; \iota = 1\, {\rm or}\, 2\; .
\label{eq:E19}
\end{equation}
Although in our model $\beta_1$ and $\beta_2$ are arbitrary parameters without
any physical meaning, their natural lower limits should not be smaller than
the corresponding thermal widths, otherwise one may be trapped into the fitting
of noise.
Then it follows that the ratio $\zeta \equiv \beta_2/\beta_1$ ranges between
$\sqrt{m_1/m_2}$ and 1 yielding $\beta^2_\iota-v^2_{\rm th,\iota} > 0$ for
any parameter sets.

\section{Practical implementation}

In the foregoing section we formulated a method to derive the kinetic temperature
by using the Fourier transform technique. 
We have so far dealt with noiseless data. Now we turn our attention to the case
of noisy data.

To calculate $T$ from equation~(\ref{eq:E16}) we need first of all to estimate
$2M+3$ parameters of model defined by equation~(\ref{eq:E13})~:
$\beta_1, \zeta, \xi, v_j, a_{1,j}$, where
$j = 1, ... , M$. 
Since we are primarily interested in the recovery of $T$, another set of the parameters
appears to be more convenient~: 
$\theta = \{T, \zeta, \xi, v_j, a_{1,j}\} \equiv \{\theta_1, ... ,\theta_{2M+3}\}$, 
with $\theta$ standing for a parameter vector. 
The standard $\chi^2$ minimization is used to estimate $\theta$~:
\begin{equation}
\chi^2(\theta) = \frac{1}{\nu} \sum^{M'}_{i=1}
\left[ \frac{I(\lambda_i) - I^\ast(\lambda_i,\theta)}{\Delta I_i} \right]^2 \; ,
\label{eq:E20}
\end{equation}
where $I(\lambda_i)$ is the simulated (or observed) normalized
intensity within the $i$th pixel of the line profile and
$I^\ast(\lambda_i,\theta) \equiv \exp\{-\tau(\lambda_i,\theta)\}$ 
is the calculated intensity. 
Both are convolved with the instrumental function $\phi_{\rm sp}$. Further, 
$\Delta I_i$ is a simulated (or experimental) error, 
$M'$ is the number of pixels involved in the sum, and 
$\nu = M' - 2M - 3$  is the number of degrees of freedom.
As already mentioned above the choice of
the number of modes $M$ is somewhat arbitrary
but the minimum number of $M$ is determined by the appropriate quality of the fit
($\chi^2 \lesssim 1$). 

Model calculations with different synthetic spectra have shown, however, that 
the minimization according to~(\ref{eq:E20}) does not allow to recover
the kinetic temperature with a sufficiently high accuracy (say 10\%).
The deviations of the recovered temperature from its underlying value
exceeded in some cases 100\%.
This failure is easy to understand~: our model for $I^\ast$ is
constructed using several smoothing operators [convolution with the
instrumental function and summing in (\ref{eq:E13})] and,
as a result, we have here a typical {\it ill-posed} problem
(see, e.g., Tikhonov \& Arsenin 1977). Parameters in ill-posed problems are
known to be very sensitive to data perturbations, which results
in the inappropriate large dispersions. 
The ill-posed problems often occur in the astronomical
image reconstructions as well, and several methods have been developed to
overcome this obstacle (see Wu 1997 and references cited therein). 
In order to stabilize the
estimation of $\theta$, we need to augment $\chi^2$ by 
a regularization term penalizing large values of parameters' errors.
Applied to our particular problem this technique is described below.

\subsection{The Entropy-Regularized $\chi^2$-Minimization procedure}

This section concerns the Entropy-Regularized $\chi^2$-Minimization [ERM]
procedure developed to estimate the best fit value for the parameter vector $\theta$.
To begin with, we formulate the ERM method. Then we state the problem in mathematical
language. Finally, we discuss some practical implementations. 

Instead of minimizing $\chi^2$ defined in equation (\ref{eq:E20}), we now seek for the minimum
of another objective function 
\begin{equation}
{\cal L}_\alpha(\theta) = \chi^2(\theta) + \alpha\,\psi(\theta)\: ,
\label{eq:E21}
\end{equation}
where $\psi$ is a penalty function and $\alpha$ the regularization parameter 
which is to be estimated.
By minimizing~(\ref{eq:E21}) we find the most probable value of the 
parameter vector $\theta_0$. 

The choice of $\psi$, especially for nonlinear problems, 
occurs rather heuristically. Since our purpose is to get the most accurate
estimation of $T$, it is reasonable to make $\psi$ penalize large error in
temperature. Having  $aT = \beta^2_1\,(1-\zeta^2)$ and 
$\Delta T/T \propto 1/\beta_1\,(1-\zeta^2)$
the easiest choice will be                
\begin{equation}
\psi = \frac{1}{\beta_1 (1 - \zeta^2)}\: ,
\label{eq:E22}
\end{equation}
which means that $\alpha$ is in units of velocity.

In spite of its simplicity this function
works quite effectively and
enables us to locate the true value of temperature very
closely. 
[It should be noted that when we change the parameter set and make $T$
be a free variable, the form of the regularization term remains the same 
but with $\beta_1 = \sqrt{aT/(1 - \zeta^2)}$.
This corresponds to
the minimization of the relative error in $\beta_1$].

The determination of the optimal value of $\alpha$ is the most crucial point
in all regularization procedures. It remains an open question still for linear
problems, although the method of Lagrange multipliers, Bayesian estimations
and maximum entropy (MaxEnt) concept proved to be very successful in many cases.
Optimality criterion applied for our problem is motivated by the idea of
minimization of the cross-entropy which is in principle equivalent to 
the MaxEnt approach (Wu 1997). In detail our procedure is described as follows.

Since entropy is a probabilistic measure of the uncertainty of the
information obtained,
we will consider the parameter vector $\theta$ as a random variable 
with the probability density function (pdf) $p_\alpha(\theta)$.
In this notation index $\alpha$ marks the pdf 
which corresponds to a given value of the regularization parameter $\alpha$.  
Although $p_\alpha(\theta)$ may have in general a skew peak, we will 
assume that a Gaussian distribution 
centered at $\theta_0$ captures most of the probability mass
of the true pdf. 
Then 
\begin{equation}
p_\alpha(\theta) = \frac{1}{\sqrt{\det(2\pi C)}}\,
\exp\{ -\frac{1}{2}\,(\theta - \theta_0)\,C^{-1}\,
(\theta - \theta_0)^t \}\: ,
\label{eq:E23}
\end{equation}
where $C$ is the error-covariance matrix.

This matrix may be estimated through the matrix of second derivatives of
${\cal L}_\alpha(\theta)$ according to its 
expansion into a Taylor series in the vicinity of the minimum~:
\begin{equation}
{\cal L}_\alpha(\theta_0 + \Delta\theta) \simeq 
{\cal L}_\alpha(\theta_0) + 
\frac{1}{2}\Delta\theta\,\frac{\partial^2 {\cal L}_\alpha(\theta_0)}
{\partial \theta^2}\,\Delta\theta^t\: ,
\label{eq:E24}
\end{equation}
and 
\begin{equation}
C_{ij} = \frac{1}{2}
\frac{\partial^2 {\cal L}_\alpha(\theta_0)}
{\partial\theta_i\partial\theta_j}\: .
\label{eq:E25}
\end{equation}

If the errors are uncorrelated, equation~(\ref{eq:E23}) simplifies to
the product of the individual parameter distributions~:
\begin{equation}
p_\alpha(\theta) = \frac{1}{\sqrt{\prod^{\nu'}_{j=1}(2\pi\sigma^2_j)}}\,
\exp\{ -\frac{1}{2}\,\sum^{\nu'}_{j=1}\,(\theta_j - \theta_{0,j})^2/\sigma^2_j\}\: ,
\label{eq:E26}
\end{equation}
where $\nu' = 2M + 3$ and $\sigma_j$ are the parameter dispersions.

To estimate exactly the level of ${\cal L}_\alpha$
that corresponds to the region containing a certain percentage
(say 68\% = $1\sigma$ for normal distribution) 
of the total probability we would need to carry
out quite extensive Monte-Carlo simulations at each step of our iterative procedure. 
However, in order to estimate roughly the parameter dispersions
we may use instead of
$\Delta{\cal L}_\alpha = {\cal L}_\alpha(\theta_0 + \Delta\theta) -
{\cal L}_\alpha(\theta_0)$
the corresponding value of $\Delta\chi^2_{\nu'}$ which is computed according to
Press {\it et al.} (1992).
Since the regularization term in ${\cal L}_\alpha$ is kept small as
compared with the $\chi^2$ term, this estimation 
of the dispersions is expected not to deviate significantly from the true value.
Then the variances in~(\ref{eq:E26}) can be evaluated as
\begin{equation}
\sigma^2_j = \frac{\Delta\chi^2_{\nu'}}{\nu}\,C^{-1}_{jj}\: .
\label{eq:E27}
\end{equation}

The definition of the cross-entropy which is 
the distance between the two distributions $p_1$ and $p_2$ 
(called also Kullback measure) is given by (cf., Csisz\'ar, 1996;  Wu, 1997)~:
\begin{equation}
{\cal K}(p_1||p_2) = \int^{\infty}_{-\infty}\,
p_1(x)\,\log\left[ p_1(x)/p_2(x)\right]\, {\rm d}x\: \geq 0\: .
\label{eq:E28}
\end{equation}
It is the non-additive, non-symmetric measure being equal to 0 only in case
$p_1 = p_2$.

Now suppose that the optimal pdf $p_{\rm opt}(\theta)$ is known. Then to
estimate the appropriate $\alpha$-value we have to minimize the distance
between $p_{\rm opt}$ and 
$p_\alpha$, i.e. the cross-entropy 
${\cal K}(p_\alpha||p_{\rm opt})$. 
Taking the pdf in the form~(\ref{eq:E26}), we obtain 
\begin{equation}
{\cal K}(p_{\alpha}||p_{\rm opt}) = \sum^{\nu'}_{j=1}\,
\log\left[\frac{\sigma^2_j(\alpha_{\rm opt})}{\sigma^2_j(\alpha)}\right] +
\frac{1}{2}\left[\frac{\sigma^2_j(\alpha)}{\sigma^2_j(\alpha_{\rm opt})} - 1\right] +
\frac{\left[\theta_{0,j}(\alpha) - \theta_{0,j}(\alpha_{\rm opt}) \right]^2}
{2 \sigma^2_j(\alpha_{\rm opt})}\: .
\label{eq:E29}
\end{equation}

In general we are not given a priori the optimal distribution, but we can
construct it adaptively from the sequence of solutions obtained for successive
values of $\alpha$ ranging from  $\alpha_{\rm min}$ to $\alpha_{\rm max}$
with a step of $\Delta\alpha$. Using the estimation
from the previous step as a current approximation for $p_{\rm opt}$ we obtain the
following iterative sequence~: 
${\cal K}(p_{\alpha_1}||p_{\alpha_{\rm min}}),\, 
{\cal K}(p_{\alpha_2}||p_{\alpha_1}),\, ...\, ,
{\cal K}(p_{\alpha_{\rm max}}||p_{\alpha_{\rm max-\Delta \alpha}})$.

Numerical experiments with synthetic spectra have shown
that this sequence really converges to 0. The optimum value
of $\alpha_{\rm opt}$, i.e. the value for which the recovered $T^\ast$ was most close to the
underlying $T$ used to produce the spectrum
was found to coincide with the point where the curvature of the trajectory 
${\cal K}(\alpha_i) \equiv {\cal K}(p_{\alpha_i}||p_{\alpha_{i-1}})$ passed
through the maximum (cf. Lucy 1994). Fig.~1 illustrates this behavior.

As known, the curvature of the continuous curve $y(x)$ is
\begin{equation}
\mu = \ddot{y}/ (\dot{y}^2 + 1)^{3/2}\: ,
\label{eq:E30}
\end{equation}
where dots indicate differentiation.
It is obvious that the value of $\mu$ depends 
on the units in which $y$ and $x$ are measured. 
Therefore
in order to get uniform results we always computed  $\mu$ for the values ${\cal K}$
and $\alpha$ normalized to 1, i.e. in the coordinates 
\begin{equation}
\hat{\cal K}, \hat{\alpha} = 
({\cal K}-{\cal K}_{\rm min})/({\cal K}_{\rm max}-{\cal K}_{\rm min})\, ,
(\alpha - \alpha_{\rm min})/(\alpha_{\rm max}-\alpha_{\rm min})\: .
\label{eq:E31}
\end{equation}

The derivatives in~(\ref{eq:E30}) are computed numerically through the common finite
differences: 
$\ddot{\cal K}_i = 
({\cal K}_{i+1} - 2{\cal K}_i + {\cal K}_{i-1})/\Delta\alpha^2$
and 
$\dot{\cal K}_i = ({\cal K}_{i+1} - {\cal K}_{i-1})/2\Delta\alpha$.
Since
numerical differentiation is rather sensitive to any even small uncertainties in
the function values, to get the smooth distribution of $\mu$ we need to
calculate the 
${\cal K}$ values with very high accuracy. This means that the minimization
of ${\cal L}$ is to be continued till the stable global minimum with 
$|\Delta {\cal L}/{\cal L}| < 10^{-6}$ is achieved.
The approximation of ${\cal K}(\alpha)$ with smoothing spline
can be used as well to obtain the stable numerical derivatives. 
But it should be noted that if ${\cal K}(\alpha)$ is computed with significant errors, then
the smoothing spline may noticeably move the maximal curvature point
to a wrong value of $\alpha_{\rm opt}$. Thus the only way to derive the
reliable estimation of $\alpha_{\rm opt}$ (and hence of $T$) is to carry out the
calculations on each step of the procedure as accurate as possible.

It should be pointed out before going ahead that we can test the estimated value
of $\alpha_{\rm opt}$ by using the intrinsic properties of the Kullback measure.
As noted above, cross-entropy is the non-symmetric measure, 
that is, ${\cal K}(p_1||p_2) \neq {\cal K}(p_2||p_1)$. This
property can be used to verify the obtained estimations
in the following way~: 
we compute ${\cal K}$-values  for both upward 
$\alpha = \alpha_{\rm min}\,(\Delta\alpha)\,\alpha_{\rm max}$ 
and downward 
$\alpha = \alpha_{\rm max}\,(-\Delta\alpha)\,\alpha_{\rm min}$ sequences 
and then estimate the corresponding optimal values of $\alpha$. 
Their coincidence for both directions may be considered as
a simple validity proof.
Another useful property 
is the non-additivity of the Kullback measure, i.e.
${\cal K}(p_{\alpha+\Delta\alpha}||p_\alpha)  \neq 
{\cal K}(p_{\alpha+\Delta\alpha/2}||p_\alpha) + 
{\cal K}(p_{\alpha+\Delta\alpha}||p_{\alpha+\Delta\alpha/2})$.
It means that if the solution of the inverse problem does exist then
the ${\cal K}(\alpha)$-sequences should yield the same $\alpha_{\rm opt}$-value
independently of the step size $\Delta\alpha$.

The limiting values of $\alpha_{\rm min}$ and $\alpha_{\rm max}$ 
should be chosen in such a way that the expected goodness-of-fit does
not deteriorate strongly ($\chi^2 \lesssim 1$). 
Although the natural choice of the minimum value seems to be
$\alpha_{\rm min} = 0$, in some cases 
the $\alpha$-sequence may be continued into the region of negative $\alpha$ 
keeping inequality ${\cal L}_\alpha > 0$ valid. The extension of the
$\alpha$-sequence below zero becomes especially essential step when the preliminary
calculations reveal a low gradient ${\cal K}(\alpha)$-curve
over the whole interval $[0,\alpha_{\rm max}]$
indicating that the maximum curvature point lies outside.

The method described makes it possible to recover the underlying kinetic temperature
from spectra of different complexity and quality. It should be 
mentioned, however, that the procedure
yields only a `best fit' value for parameter(s) without confidence intervals
(a point-like estimation) since the dispersions
obtained can be over- or underestimated due to the reasons considered above. The
actual error limits may be revealed only through the numerical experiments with synthetic
spectra. 
The model calculations with noisy data (SNR = 100) and spectral resolutions (FWHM) ranging
from 3~km~s$^{-1}$ to 12~km~s$^{-1}$
have shown, however, that similar unsaturated profiles of  
 C$^+$, Si$^+$ and Fe$^+$ ions provide 
the accuracy of the kinetic temperature estimation as high as 10\%.

\subsection{Numerical experiments}

In the previous section we have considered the ERM algorithm.
The remaining problem now is to show the accuracy of the ERM solutions
and hence their reality.

Since our computational procedure contains several {\it ad hoc}
assumptions which can hardly be backed with 
unambiguous mathematical justification,
it is natural to prove its validity through the recovering
the kinetic temperature from the synthetic absorption-line spectra at first. 
For this purpose we have generated a variety of line profiles employing
different sets of physical parameters and several random 
realizations of the one-dimensional
velocity fields. The calculations were performed for the pairs of lines
C~II$\lambda1334$~+~Si~II$\lambda1260$ 
and C~II$\lambda1334$~+~Fe~II$\lambda2600$  which are frequently observed in
interstellar and quasar absorption spectra. The procedure may be applied to any
other suitable pairs as well.

The physical parameters specifying 
our cloud model are the column densities of the elements in a pair,
the dispersion of the turbulent 
velocity field $\sigma_t$, and the kinetic temperature $T$. 
Besides, all synthetic
spectra have been convolved with the instrumental profile of different FWHM.
In order to make the synthetic spectra look like true observations with an SNR of 100,
corresponding Gaussian noise has been added.
The model parameters are listed in Table~1.

To calculate synthetic spectra we need to generate stochastic velocity fields
at first.
Modeling of the stationary random velocity fields $v(s)$ 
with different correlation ranges
in space is performed through the moving average method described in Appendix A.
To generate the correlated fields,
the following parameterized family of correlation functions was chosen
(see Elliott {\it et al.} 1997)~: 
\begin{equation}
{\cal C}_{\ell,\varepsilon}(s) = \left( 1 + \frac{s^2}{\ell^2} \right)^{\varepsilon/2-1}\,
\cos\left[ (2-\varepsilon)\,\arctan\left(\frac{|s|}{\ell}\right)\right]\: ,
\label{eq:E32}
\end{equation}
where $\ell$ is a linear scale proportional to the distance over which
there is appreciable correlation between the velocity values at two points
(see Monin \& Yaglom 1975).
This family is widely used in the turbulent flow modeling and 
provides a diversity of scaling behavior as the parameter $\varepsilon$ 
varies in the range $-\infty < \varepsilon < 2$. 
For instance, regions of negative correlations for the velocity field may be obtained 
with $\varepsilon < 1$, whereas for $\varepsilon \geq 1$ we have the regime of 
positive correlations with very long ranges. Parameters $\ell$ and
$\varepsilon$ used to generate the velocity fields are listed in Table~1 as well
($\ell$ is given in units of $L$).

Listed in Table~1, models~1--11 were constructed
to settle the accuracy of the kinetic temperature measurement at different spectral
resolutions. The synthetic spectra were calculated for the fixed values of
$T = 15000$ K\, and $\sigma_{\rm t} = 15$~km~s$^{-1}$, using different velocity fields.
The velocity fields were chosen rather at random because our aim was mainly to
show the ability of the method to recover the correct temperature from almost any
velocity fields. 
Thereafter the spectra were convolved with Gaussian
functions with FWHM = 3, 6, 7, 9, and 12~km~s$^{-1}$. The spectral resolution
of 3~km~s$^{-1}$ was achieved in the studies of the interstellar absorption lines
with HST, whereas the intergalactic spectra are being observed at present with the
resolutions 6--12~km~s$^{-1}$. The ratio ${\rm FWHM}/v_{\rm th,2}$ 
in our model series lies within
the range 1--4 meaning that the smoothing effect for ${\rm FWHM} \geq 6$~km~s$^{-1}$
becomes very pronounced.

To apply our ERM procedure to another kind of absorption spectra,
we have calculated the synthetic spectra for 
$\sigma_{\rm t} = 5$~km~s$^{-1}$
and the temperatures  $T = 15000$ K\, and 2000 K\, (model~12 and 13, respectively).  
The last two models~14 and 15 were constructed to test the procedure for different
ratios of  $m_1/m_2$ and $\sigma_{\rm t}/v_{\rm th,2}$.

Table~2 lists the recovered kinetic temperatures $T^\ast$ along with some other
estimated parameters. The comparison of columns (7) and (8) shows the high efficiency
of the proposed computational procedure which allows to estimate $T$ with an
accuracy of $|\Delta T/T| \lesssim 10$\% even for the rather smeared out 
(models~7--11) spectra. 

In column (6) of Table~2 we list the values of $\zeta = \beta_2/\beta_1$. Numerical
experiments have shown that for $\zeta \gtrsim 0.9$
the dispersion of $T^\ast$,\, $\sigma_{{\rm T}^\ast}$, calculated through~(\ref{eq:E27}) 
changes its behavior and begins to increase with increasing the regularization
parameter $\alpha$. Such peculiar pattern of $\sigma_{{\rm T}^\ast}$ stems from the
fact that at $\zeta \gtrsim 0.9$ the off-diagonal elements of the error-covariance matrix $C$
in (\ref{eq:E23}) are already not small as compared with $C_{ii}$ and, hence, the
approximation~(\ref{eq:E26}) becomes rather vague since the errors become strongly
correlated. However, the iterative procedure continues to deliver an acceptable
$T$ estimation. The procedure fails when $\zeta \approx 1$. 
This is the case when the ratio ${\rm FWHM}/v_{\rm th,2}$ is large and
the intensity fluctuations within the line profiles
are considerably smoothed which in turn makes the widths of the subcomponents
practically indistinguishable.

The consequent steps of our computational scheme are illustrated in Figures~2--4.
The case of positive correlations for the large scale velocity field
($\varepsilon = 1.1$ and 1.0) is presented in Fig.~2 and 4, respectively.
In Fig.~2, panel (a) shows
with the solid line the adopted correlation function ${\cal C}_{\ell,\epsilon}$,
and with the dotted one -- the  autocorrelation function (ACF) 
calculated for the random realization of $v(s)$.
The computed ACF may deviate from the exact correlation function 
because of the finite thickness of the absorbing region. 
Panel (b) shows the corresponding velocity field $u(x) = v(x)/\sigma_{\rm t}$,
where $x = s/L$. This field leads in turn to the complex profiles of C~II and Si~II.
Both profiles are convolved with a Gaussian function having the width of
3~km~s$^{-1}$ (FWHM), and then 
a Gaussian noise of S/N = 100 has been added. The final
patterns are shown in panels (c) and (d) by dots with corresponding error bars.
The C~II and Si~II lines
were fitted simultaneously employing (\ref{eq:E13}), (\ref{eq:E20}) and (\ref{eq:E21}).
The fit was performed with the increasing number of modes $M$ until 
a $\chi^2$-value of $\lesssim 1$ was achieved.
In panels (c) and (d) the best-fitting profiles with $M = 7$ equidispersion 
components (located at the velocities shown by the tick marks)
are plotted as solid curves superimposed on the simulated data. 
The joint $\chi^2$-value per degree of freedom ($\nu$ = 129) 
is indicated at the bottom of panel~(d). 
Panel (e) illustrates that if a feasible MaxEnt solution does exist, then
the Kullback measure $\hat{\cal K}(\hat{\alpha})$ calculated through 
(\ref{eq:E29},\ref{eq:E31})
converges with increasing $\hat{\alpha}$. The optimal value of $\alpha = 0.0$
corresponding to the maximum curvature of the trajectory $\hat{\cal K}(\hat{\alpha})$ 
leads to the estimation $T^\ast = 15097$ K. Given the kinetic temperature, we may 
calculate the radial-velocity distribution function 
$\phi_{\rm t}(v)$ using equation~(\ref{eq:E19}) and either C~II or Si~II spectrum.
For both sets of data the solution is expected to be identical. The results obtained
are shown in panel (f) by  solid curves.
The histogram plotted in panel (f)
is the underlying distribution function 
$p(v)$ which corresponds to the velocity 
profile $v(x) = \sigma_{\rm t} u(x)$ shown in panel~(b).
Since all curves are very much alike, we consider this result as an  additional
justification of our procedure.
 
The regime of negative correlations $(\varepsilon = -0.5)$ for the random velocity
field is illustrated in Fig.~3 (here FWHM = 7~km~s$^{-1}$). The case for the lowest
resolution of 12~km~s$^{-1}$ is demonstrated in Fig.~4. The fitting parameters for these
examples are listed in Table~2. Analyzing all results, we may conclude that
the procedure yields stable estimations of $T$ and $\phi_{\rm t}(v)$ for
the whole range of spectral resolutions considered.

\section{Discussion}

\subsection{Comparison with the standard Voigt deconvolution}

Based on the Voigt fitting, most previous approaches extracted information
on the gas kinetic temperature from the values of $b$ found for the various
subcomponents of the complex profiles. As a result, absorbing regions have
usually been characterized by a variety of $T$ values which, in some cases,
ranges from $\approx 0$~K up to $\approx 10000$~K (e.g., Spitzer \& Fitzpatrick 1995).
It seems quite natural to apply the standard Voigt fitting procedure to our
synthetic spectra to compare its result with the estimation obtained by
the proposed method. 

To carry out such analysis, we generated the C~II$\lambda1334$ and 
Fe~II$\lambda2600$ absorption profiles (Fig.~5, model~14 from Table~1)
convolved with the instrumental function of FWHM~=~3 km~s$^{-1}$.
The corresponding Gaussian noise with an S/N of 100 was added as well.
Then we considered the spectra shown in Fig.~5 as `observational' data
and tried to analyze them in the standard manner, i.e. to fit them
with Voigt profiles until a satisfactory result with $\chi^2 \lesssim 1$ was
achieved. A simultaneous fit to the 10 subcomponents gave the reduced
$\chi^2$ per degree of freedom of 0.687 $(\nu = 172)$.

The parameter estimates are listed in Table~3 along with the corresponding
values of the gas temperatures $T^\ast_i$ and the turbulent velocities
$\sigma^\ast_i$ for the `individual cloudlets'. The revealed values clearly
demonstrate a wide spread of the apparent $b$-parameters resulting in
the large variations of $T$~: from $\approx 10$~K to $\approx 18000$~K 
whereas the spectra have been produced by using a single underlying temperature
$T = 10000$~K. Moreover, taken at face values, the relative widths of
$b_i({\rm C~II})$ and  $b_i({\rm Fe~II})$ lead to unlikely small values of
$\sigma^\ast_i \approx 1 - 4$~km~s$^{-1}$, 
i.e. to a significantly lower turbulent velocity than adopted 
$\sigma_{\rm t} = 20$ km~s$^{-1}$.

The same synthetic spectra having been treated by the proposed procedure
yield a single value for the gas temperature of 10461~K (see Table~2).
The corresponding equidispersion deconvolution (with $M = 11$ components)
is plotted in panels $c$ and $d$ in Fig.~5.

These results lead to the conclusion that the spread of the gas temperature
found in some regions of the ISM and in the extragalactic intervening clouds
may be unphysical and stems purely from an inappropriate method
of data analysis.

\subsection{The C~II and Si~II profiles observed in QSO~1937--1009}

We turn now to a brief discussion of the application of 
the present computational technique to recover the
kinetic temperature from observational data. For this purpose we
have chosen the high quality Keck spectra of the quasar Q~1937--1009
(Tytler {\it et al.} 1996) showing metal absorption lines at the
redshift $z = 3.572$. Since this system is the only one where the
deuterium absorption was observed at high redshift
in both the D~Ly$\alpha$ and Ly$\beta$ lines, it has been thoroughly
studied using different methods (Tytler \& Burles 1997;
Levshakov {\it et al.} 1998a; Burles \& Tytler 1998). 

According to Tytler \& Burles, all metal absorption lines seen at this redshift
(C~II, C~IV, Si~II, Si~IV) `appear to have the same profiles'. 
This may lead to the suggestion that these metals are similarly distributed
in the velocity space and hence a single gas temperature could be
sufficient to describe their profiles. To estimate its value,
we have selected the C~II$\lambda1334$ and Si~II$\lambda1260$ lines observed
with FWHM~= 8~km~s$^{-1}$ and S/N~= 56 and 100, respectively.  
The original normalized data are shown in Fig.~6 by dots with the corresponding
error bars.

The Voigt fitting analysis
previously performed by Tytler {\it et al.} (1996) revealed
two gas clouds separated by 15~km~s$^{-1}$ (the {\it blue} and the {\it red}
components). The physical parameters found for these components are the
following~ (Tytler \& Burles)~:
$T_{\rm blue}~= 17100 \pm 1300$~K, $\sigma_{\rm blue}~= 3.35 \pm 0.56$~km~s$^{-1}$, 
$T_{\rm red}~= 19000 \pm 5800$~K, $\sigma_{\rm red}~= 6.75 \pm 0.28$~km~s$^{-1}$
(where $\sigma \equiv b_{\rm turb}/\sqrt{2}$).

This system has also been analyzed by different method based on the
mesoturbulent concept (Levshakov {\it et al.} 1998a). In this study the
H+D~Ly$\alpha$, Ly$\beta$ and higher order Lyman series absorption lines 
(up to Ly-19) have been treated using
the reverse Monte-Carlo (RMC) procedure (computational details are given
in Levshakov, Kegel \& Takahara 1998a,b). The gas temperature, $T^\ast$, and
the ensemble turbulent velocity dispersion, $\tilde{\sigma}^\ast_{\rm t}$,
estimated within the framework of the one-cloud model were found to lie
in the ranges from 10900 to 14300~K and from 18.5 to 22.3~km~s$^{-1}$,
respectively.

Applying the present procedure to the C~II and Si~II lines, we obtained
the gas temperature $T^\ast = 14090$~K. Table~4 and Fig.~6 illustrate our results.
The best-fitting profiles with $M = 5$ equidispersion components are
shown in panels $a$ and $b$ by solid lines with the tick marks indicating
the velocity positions for the separate components. Fig.~6c demonstrates the
radial-velocity distribution function (solid line curve) calculated
through equation~(\ref{eq:E19}). 
The square root of the central second moment of this 
distribution $\sigma^\ast_{\rm t}~= 8.6$~km~s$^{-1}$ depicted in panel $c$
may be considered as a measure of the one-dimensional velocity scatter
along the space coordinate. 
This quantity naturally falls below the ensemble velocity dispersion
$\tilde{\sigma}^\ast_{\rm t}$ because the sample of the velocity
components parallel to the line of sight is not complete in the statistical sense
(see Levshakov \& Kegel 1997, for details).
Also shown in Fig.~6c is a histogram which represents the velocity distribution
derived from the analysis of the hydrogen and deuterium lines by the RMC procedure. 

In general we would expect \ion{H}{1} and metals to give similar temperatures.
They would be identical if absorbing gas was fully homogeneous.
Analyzing our results on the $z = 3.572$ system we may
conclude that both RMC and ERM methods give the consistent estimate of the
kinetic temperature, $T^\ast \simeq 14000$~K.
As for the
velocity distribution of neutral hydrogen and deuterium  and the velocity
distribution of ionized metals, they apparently differ (see Fig.~6c). 
This difference, however, stems from the observed asymmetry of the deuterium and
the optically thin metal absorption lines (Tytler \& Burles)~:
the former exhibits the blue-side asymmetry, whereas the latter show the
red-side skewness. The interpretation of this result may be as follows.
It is generally believed
that most metal absorption-line systems observed in QSO spectra
arise in the far outer parts of intervening galaxies or in the 
intergalactic medium where
the main source of ionization is photoionization due to the background
UV-flux. If the temperature of about 14000~K is the
mean value for the absorbing gas at $z = 3.572$
which is in thermal equilibrium, then variations in the fractional ionization
of different species along
the line of sight are mainly governed by
the density fluctuations $n_{\rm gas}$ within this region because the thermal
and ionization state of the gas is described by the ionization parameter $U$ which
is the ratio of the density of ionizing photons to the total gas density.
Ionization models of optically thin in the Lyman continuum intergalactic
gas at constant pressure, photoionized by QSOs, show that the equilibrium temperature
is weakly dependent on $U$ for metallicities below solar and 
$-4.7 < \log U < -1.8$, where $T \propto U^{1/8}$ (Donahue \& Shull 1991).
The deviations between the apparent radial-velocity distribution functions
may occur when the fractional ionization of different elements
responds in different ways to the changes in $U$. 
Such deviations may be utilized to place constraints
on the density fluctuations within the absorbing gas.
This problem will be studied in detail elsewhere.

\section{Conclusions}

The outcome of our study, aiming to provide a measurement of
the gas kinetic temperature from {\it similar} 
complex absorption-line spectra of
ions with different atomic weights is the following.

\medskip\noindent
(1) A new method is developed for using interstellar and intergalactic
absorption-line spectra to investigate the intervening gas temperature.
It is based on the assumption that a pair of elements has the same
temperature and bulk motion within the absorbing region resulting in
the similar observed spectra.
The method utilizes the Fourier transform of the observed profiles 
with the subsequent Entropy-Regularized Minimization
of the $\chi^2$ function augmented by a penalty function to gain a stable
solution of the inverse problem of spectral analysis [the ERM method].

\medskip\noindent
(2) The computational procedure has been tested on a variety of synthetic
spectra being generated in stochastic velocity fields with different
correlation scales.  
The procedure proved to be very effective and robust 
allowing to recover the temperature
with high accuracy, $|\Delta T / T| \lesssim 10$~\%, in the range of 
spectral resolutions from 3~km~s$^{-1}$ to 12~km~s$^{-1}$ (FWHM). 

\medskip\noindent
(3) The application of the standard Voigt fitting
analysis to the synthetic spectra calculated in the framework of the
homogeneous temperature model with $T = 10^4$~K revealed a wide scatter    
of the gas temperature among the individual `cloudlets', -- 
from 10~K to $1.8\times10^4$~K. 
This result demonstrates that the scattering of $T$ 
reported in the literature may have non-physical nature and
may be caused by an inappropriate methodology. 

\medskip\noindent
(4) The asymmetric C~II$\lambda1334$ and Si~II$\lambda1260$ lines observed 
with the Keck~I telescope at the redshift $z = 3.572$ towards 
quasar  1937--1009 by Tytler {\it et al.} have been treated by the proposed method
in order to evaluate the underlying gas temperature. The obtained estimation of
$T = 14090$~K lies in the range consistent with the previously derived values
of $T$ from the analysis of the H+D Ly$\alpha$ and Ly$\beta$ and higher order
Lyman series profiles 
by the reverse Monte-Carlo method (Levshakov {\it et al.} 1998a).

\bigskip\noindent
It should be emphasized once again that the proposed method can be applied
to the similar spectra only. 
In reality different (non-similar) profiles are observed as well (see, e.g.,
Huang {\it et al.} 1995, or Songaila 1998). 
They may be caused by a variety of physical reasons, -- overlapping of
cold and hot gas clouds due to the Doppler shifts of their radial velocities,
variations in the ionizing flux produced by local sources and/or collisional
ionizations in higher density sub-regions  both leading to the diversity in the 
ionization fractions for different elements, etc., -- and hence there cannot
be a single physical model to treat such observational data.
 
\acknowledgments

The authors are grateful to David Tytler 
for sharing his calibrated Keck~I/HIRES echelle spectra of Q1937--1009
and for a careful reading the manuscript.
We also thank Wilhelm Kegel for his valuable comments and suggestions
which help us much in this study. 
S.A.L. gratefully acknowledges the hospitality of the 
Osaka University and the National
Astronomical Observatory of Japan where this work was performed. 

\appendix

\section{The moving average method for simulating Gaussian random fields}

Gaussian random fields arise in a wide variety of astrophysical problems
including gas flows, large scale motions, turbulence etc.
To study these problems one needs to model the stochastic fields with given
characteristics (power spectrum or covariance function).
The moving average method employed in our modeling the intervening velocity
fields is able to reproduce the true covariances over the larger scales as
compared with the most common spectral (Fourier) approximation (Pen 1997;
Elliott {\it et al.} 1997). The description below follows in
general that presented in Elliott {\it et al.}.

Consider the Gaussian random field $v(s)$ with the covariance function $R(s)$ and
the power spectrum $S(k)$. $R(s)$ and $S(k)$ are related through common equations
representing the direct and inverse Fourier transforms correspondingly~:
\begin{equation}
S(k) = \int^{+\infty}_{-\infty} R(s)\,{\rm e}^{2\pi iks}\,{\rm d}s \; ,
\label{eq:A1}
\end{equation}
and
\begin{equation}
R(s) = \int^{+\infty}_{-\infty} S(k)\,{\rm e}^{- 2\pi iks}\,{\rm d}k \; .
\label{eq:A2}
\end{equation}

Define now the function $G(s)$ which is the Fourier transform of the square root
of the power spectrum $S(k)$~:
\begin{equation}
G(s) = \int^{+\infty}_{-\infty} S^{\case{1}{2}}(k)\,{\rm e}^{2\pi iks}\,{\rm d}k \; .
\label{eq:A3}
\end{equation}

Then $v(s)$ can be represented in the form~:
\begin{equation}
v(s) = \int^{+\infty}_{-\infty} G(s' - s)\,{\rm d}W(s')\; ,
\label{eq:A4}
\end{equation}
where $W(s)$ is the Wiener process with unit variance 
(see, e.g., Cox \& Miller 1965).

The moving average method stems from the discretization of~(\ref{eq:A4}),
resulting in the approximation 
\begin{equation}
\hat{v}(s) = \sum^{p}_{i=-p}\,G(r_i - s)\,Z(r_i)\; ,
\label{eq:A6}
\end{equation}
where $Z$ is a Gaussian random variable with mean zero  and variance
$\langle Z^2 \rangle = \Delta r$, and  $\Delta r$ being the step of 
the equally spaced grid points.

Gaussian fields are considered to be equal if they have the same mean
and covariance function. Therefore to test whether the approximation~(\ref{eq:A6})
represents the real field, we have to compare the covariance function of the approximated
field with that which was adopted. In this way we may estimate the appropriate step size
$\Delta r$ and the number of the grid points $p$. The proper values of $\Delta r$ 
and $p$ should provide  minimal difference between the calculated and adopted covariance
functions.

\clearpage

\figcaption[]{
Convergence behavior of the cross-entropy defined
by~(\ref{eq:E29}). The filled circles connected by dotted line
show the normalized values of $\hat{\cal K}$ calculated in accord with
(\ref{eq:E31}). Also plotted is the curvature $\mu/\mu_{\rm max}$
(filled circles connected by dashed line) of the trajectory
$\hat{\cal K}(\hat{\alpha})$. The optimal value of
$\alpha_{\rm opt} = 0.15$ corresponds to the maximum curvature
at point $\hat{\alpha} = 0.167$. This example is calculated for
model~3 from Table~1. \label{fig1}}
 
\figcaption[]{
($a$) Autocorrelation functions~: solid curve -- adopted model,
dotted curve -- 
calculated from the random realization of the velocity field shown in panel ($b$).
($c,d$) Velocity plots of the simulated spectra (dots and 1$\sigma$ error bars)
and the calculated profiles convolved with the instrumental resolution of
FWHM = 3~km~s$^{-1}$ (solid curves), corresponding to model 2 in Table~1.
($e$) Cross-entropy as function of the normalized regularization parameter 
$\hat{\alpha}$ (filled circles connected by dotted line) and the
curvature of its trajectory (filled circles connected by dashed line).
($f$) Simulated one-dimensional velocity distribution (histogram) and
its estimation through the C~II (or Si~II) profiles. 
See text for more details.
\label{fig2}
}

\figcaption[]{
As Fig.~2, but for different velocity field and
FWHM = 7~km~s$^{-1}$ (model~4 in Table~1).
\label{fig3}
}
\figcaption[]{
As Fig.~2, but for different velocity field and
FWHM = 12~km~s$^{-1}$  (model~11 in Table~1).
The original (unconvolved) spectra
are shown in panels ($c$) and ($d$) by dotted curves as well.
Note the
doublet-like structure at $\Delta v = 40$ ~km~s$^{-1}$ in the Si~II spectrum 
which is completely smeared out after convolution.
\label{fig4}
}

\figcaption[]{
$(a,b)$ Velocity plots of the 
C~II$\lambda1334$ and Fe~II$\lambda2600$ synthetic spectra
(dots with $1\sigma$ error bars, model~14 in Table~1). 
The smooth curves show the Voigt profiles convolved with
the instrumental resolution of 3~km~s$^{-1}$ (FWHM) which produce the best
simultaneous fit to all data. The velocity positions for each of ten absorption
components are labeled by the tick marks at the top of each panel.
$(c,d)$ Same as  $(a,b)$, but for the ERM result
based on the equidispersion deconvolution of the C~II and Fe~II profiles
with a single kinetic temperature (see text for more details).
\label{fig5}
}

\figcaption[]{
$(a,b)$ Normalized data points with $1\sigma$ error bars
for HIRES echelle spectrograph observations (FWHM~= 8~km~s$^{-1}$) on the
Keck telescope of the C~II$\lambda1334$ and Si~II$\lambda1260$   
absorption features (the signal-to-noise per pixel is, respectively, 56 and 100)
at the redshift $z = 3.572$ towards the quasar 1937--1009 (Tytler \& Burles 1997).
The ERM result 
(reduced $\chi^2 = 0.331$ with 24 degrees of freedom)
for C~II and Si~II (solid lines)
is based on simultaneously fitting the multiple equidispersion lines located
at the velocities shown by the tick marks at the top of each panel. The
evaluated kinetic temperature $T^\ast = 14090$~K is plotted in panel $a$.
$(c)$ The radial-velocity distribution function $\phi_{\rm t}(v)$
restored through the observed profile of the C~II (or Si~II) line
using eq.~(\ref{eq:E19}) and the model parameters
from Table~4. The square root of the central second moment 
$\sigma^\ast_{\rm t}$ of the $\phi_{\rm t}(v)$ distribution
is also shown in the panel. 
For comparison, histogram reproduces the velocity distribution function
estimated by the reverse Monte-Carlo [RMC] procedure  (Levshakov {\it et al.} 1998a)
through the analysis of the H+D Ly$\alpha$, Ly$\beta$ and the higher order Lyman series lines
(up to Ly-19) from the same absorption-line system.
\label{fig6}
}

\clearpage

\begin{deluxetable}{cccccccc}
\footnotesize
\tablecaption{
Model parameters used to generate synthetic spectra of
C~II$\lambda1334$ and Si~II$\lambda1260$ (models~1--13)
and synthetic spectra of C~II$\lambda1334$ and Fe~II$\lambda2600$ (models~14,15).
The parameterized family of velocity correlation functions
(\ref{eq:E32}) is defined by the parameters $\ell$ and $\varepsilon$
(see text).
\label{tab1}
}
\tablewidth{16cm}
\tablehead{
\colhead{model} & \colhead{pair} & \colhead{$N$,} 
& \colhead{$\sigma_{\rm t}$,} & \colhead{$T$,} & \colhead{FWHM,} 
& \colhead{$\ell$} & \colhead{$\varepsilon$} \nl
\colhead{ } & \colhead{ } & \colhead{cm$^{-2}$} 
& \colhead{km s$^{-1}$} & \colhead{K} & \colhead{km s$^{-1}$} 
& \colhead{ } &  \colhead{ }
}
\startdata
1 & C~II & 4(13) & 15 & 1.5(4) & 3 & 0.1 & 1.0 \nl
  & Si~II& 4(12) & & & & &  \nl
2 & C~II & 4(13) & 15 & 1.5(4) & 3 & 0.1 & 1.1 \nl
  & Si~II& 4(12) & & & & &  \nl
3 & C~II & 4(13) & 15 & 1.5(4) & 6 & 0.1 & 1.1 \nl
  & Si~II& 4(12) & & & & &  \nl
4 & C~II & 4(13) & 15 & 1.5(4) & 7 & 0.25 & $-0.5$ \nl
  & Si~II& 4(12) & & & & &  \nl
5 & C~II & 4(13) & 15 & 1.5(4) & 7 & 0.25 & 0.5 \nl
  & Si~II& 4(12) & & & & &  \nl
6 & C~II & 4(13) & 15 & 1.5(4) & 7 & 0.25 & $-1.5$ \nl
  & Si~II& 4(12) & & & & &  \nl
7 & C~II & 4(13) & 15 & 1.5(4) & 9 & 0.1 & 1.0 \nl
  & Si~II& 4(12) & & & & &  \nl
8 & C~II & 4(13) & 15 & 1.5(4) & 9 & 0.1 & 1.1 \nl
  & Si~II& 4(12) & & & & &  \nl
9 & C~II & 4(13) & 15 & 1.5(4) & 9 & 0.1 & $-1.0$ \nl
  & Si~II& 4(12) & & & & &  \nl
10& C~II & 4(13) & 15 & 1.5(4) & 12 & 0.1 & 1.1 \nl
  & Si~II& 4(12) & & & & &  \nl
11& C~II & 4(13) & 15 & 1.5(4) & 12 & 0.1 & 1.0 \nl
  & Si~II& 4(12) & & & & &  \nl
12& C~II & 4(13) & 5 & 1.5(4) & 3 & 0.5 & $-1.5$ \nl
  & Si~II& 4(12) & & & & &  \nl
13& C~II & 4(13) & 5 & 2.0(3) & 3 & 0.05 & $-1.5$ \nl
  & Si~II& 4(12) & & & & &  \nl
14& C~II & 5.5(13) & 20 & 1.0(4) & 3 & 0.3 & $-1.5$ \nl
  & Fe~II& 5.3(12) & & & & &  \nl
15& C~II & 5.5(13) & 20 & 1.0(4) & 6 & 0.3 & $-1.5$ \nl
  & Fe~II& 5.3(12) & & & & &  \nl

\enddata
\tablenotetext{ }{ }
\end{deluxetable}

\clearpage

\begin{deluxetable}{cllcccrr}
\footnotesize
\tablecaption{
Model parameters derived from the synthetic spectra
through the equidispersion deconvolution (\ref{eq:E13})
with the subsequent entropy-regularized minimization of
the objective function (\ref{eq:E21}).
\label{tab2}}
\tablewidth{16cm}
\tablehead{
\colhead{model} & \colhead{$\alpha_{\rm min}$} 
& \colhead{$\alpha_{\rm opt}$} & \colhead{$M$} 
& \colhead{$\chi^2$} & \colhead{$\zeta$} 
& \colhead{{$T^\ast$\tablenotemark{a}}, K}
& \colhead{$T$, K} 
}
\startdata
1 & \hspace{0.33cm}0.0 & \hspace{0.33cm}0.2 & 8 & 0.713 & 0.86 & 14271 & 15000 \nl
2 & $-0.4$ & \hspace{0.33cm}0.0 & 7 & 0.674 & 0.77 & 15097 & 15000 \nl
3 & \hspace{0.33cm}0.0 & \hspace{0.33cm}0.15 & 4 & 0.725 
& 0.89\tablenotemark{b} & 13775 & 15000 \nl
4 & $-0.15$ & \hspace{0.33cm}0.0 & 4 & 0.364 & 0.84 & 14576 & 15000\nl
5 & \hspace{0.33cm}0.0 & \hspace{0.33cm}0.275 & 4 & 0.797 & 0.85 & 15401 & 15000 \nl
6 & \hspace{0.33cm}0.0 & \hspace{0.33cm}0.2 & 5 & 0.814 & 0.86 & 15866 & 15000 \nl
7 & $-0.037$ & $-0.032$ & 4 & 0.710 
& 0.93\tablenotemark{b} & 14530 & 15000 \nl
8 & $-0.069$ & $-0.062$ & 4 & 0.409 & 0.89 & 13537 & 15000 \nl
9 & \hspace{0.33cm}0.0 & \hspace{0.33cm}0.012 & 4 & 0.741 
& 0.95\tablenotemark{b} & 13910 & 15000 \nl
10& $-0.006$ & $-0.003$ & 2 & 0.408 
& 0.96\tablenotemark{b} & 14473 & 15000 \nl
11& $-0.0203$ & $-0.0198$ & 4 & 0.458 
& 0.93\tablenotemark{b} & 14632 & 15000 \nl
12& $-0.2$ & \hspace{0.33cm}0.5 & 2 & 0.257 & 0.71 & 16112 & 15000 \nl
13& \hspace{0.33cm}0.0 & \hspace{0.33cm}0.02 & 5 & 0.752 
& 0.92\tablenotemark{b} & 1816 & 2000 \nl
14& \hspace{0.33cm}0.0 & \hspace{0.33cm}0.3 & 11 & 0.763 & 0.75 & 10461 & 10000 \nl
15& $-0.04$ & \hspace{0.33cm}0.02 & 8 & 0.756 & 0.85 & 10081 & 10000 \nl

\enddata

\tablenotetext{a}{
$T^\ast$ -- estimated temperature, $T$ -- adopted temperature.}
\tablenotetext{b}{ increasing dispersion $\sigma_{{\rm T}^\ast}$
(see text for details).} 

\end{deluxetable}

\clearpage

\begin{deluxetable}{clcccc}
\footnotesize
\tablecaption{
The standard Voigt fitting procedure results.
Model parameters evaluated simultaneously from
the C~II$\lambda1334$ and Fe~II$\lambda2600$ synthetic spectra
(Fig.~5) by fitting ten individual subcomponents numbered in the order
of increasing relative velocity $\Delta v$. Reduced $\chi^2$
(with 172 degrees of freedom) is equal to 0.687. The adopted values for
the kinetic temperature, $T$, and the rms turbulent velocity 
dispersion, $\sigma_{\rm t}$, are, respectively,
equal to $10^4$~K and 20~km~s$^{-1}$ (model~14 in Table~1).
\label{tab3}}
\tablewidth{16cm}
\tablehead{
No. & \colhead{$\Delta v$,} & \colhead{$b_i$(C~II),} 
& \colhead{$b_i$(Fe~II),} & \colhead{$T^\ast_i$,} & \colhead{$\sigma^\ast_i$,} \nl
\colhead{ }  & \colhead{km~s$^{-1}$} 
& \colhead{km~s$^{-1}$} & \colhead{km~s$^{-1}$} & \colhead{K} 
& \colhead{km~s$^{-1}$} \nl 
}
\startdata
1 & \hspace{0.0cm}$-47.86$ & 3.75 & 2.09 &\hspace{0.2cm}8896 & 0.9 \nl
2 & \hspace{0.0cm}$-37.79$ & 6.46 & 4.90 & 16212 & 3.1 \nl
3 &\hspace{0.0cm}$-27.42$&\hspace{0.2cm}5.998&\hspace{0.2cm}5.997&\hspace{0.65cm}11&4.2 \nl
4 & \hspace{0.0cm}$-12.68$ & 5.79  & 3.64 & 18590 & 2.0 \nl
5 & \hspace{0.2cm}$-5.90$  & 5.06  & 2.50 & 17821 & 0.7 \nl
6 & \hspace{0.52cm}0.52   & 5.20  & 3.76 & 11854 & 2.3 \nl
7 & \hspace{0.52cm}9.78   & 6.07  & 4.90 & 11837 & 3.2 \nl
8 & \hspace{0.34cm}23.24  & 6.00  & 5.37 & \hspace{0.2cm}6564  & 3.7 \nl
9 & \hspace{0.34cm}33.93  & 5.86  & 5.63 &\hspace{0.2cm}2467  & 3.9 \nl
10 &\hspace{0.34cm}41.90  & 4.38  & 3.18 &\hspace{0.2cm}8310  & 1.9 \nl

\enddata
\end{deluxetable}

\begin{deluxetable}{clcc}
\footnotesize
\tablecaption{
The fit based on the entropy-regularized minimization.
Model parameters evaluated simultaneously from
the observed C~II$\lambda1334$ and Si~II$\lambda1260$
spectra (Fig.~6) by fitting five equidispersion components numbered in the order
of increasing velocity position $v_j$. Reduced $\chi^2$
(with 24 degrees of freedom) is equal to 0.331. The estimated value for
the kinetic temperature is $T^\ast = 14090$~K, and the widths of the
components are equal to $\beta^\ast({\rm C~II}) = 5.735$~km~s$^{-1}$ and
$\beta^\ast({\rm Si~II}) = 4.671$~km~s$^{-1}$ (see equation~(\ref{eq:E13})).
\label{tab4}}
\tablewidth{16cm}
\tablehead{
\colhead{No.} & \colhead{$v_j$,~km~s$^{-1}$} & \colhead{$a_j$(C~II)} 
& \colhead{$a_j$(Si~II)} \nl
}
\startdata
1 & \hspace{0.0cm}$-15.18$ & 0.813 & 0.833 \nl
2 & \hspace{0.2cm}$-7.56$  & 1.619  & 1.658 \nl
3 & \hspace{0.2cm}$-0.93$  & 2.998  & 3.071  \nl
4 & \hspace{0.52cm}8.57   & 4.930  & 5.051  \nl
5 & \hspace{0.34cm}17.62  & 0.445  & 0.456 \nl

\enddata
 \end{deluxetable}

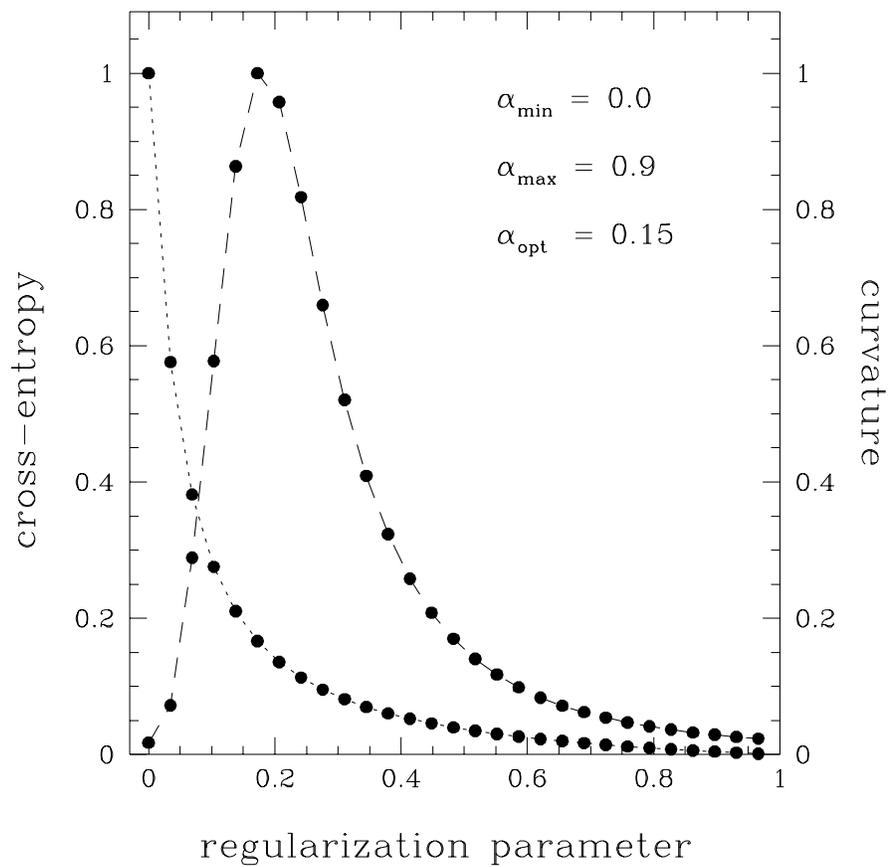
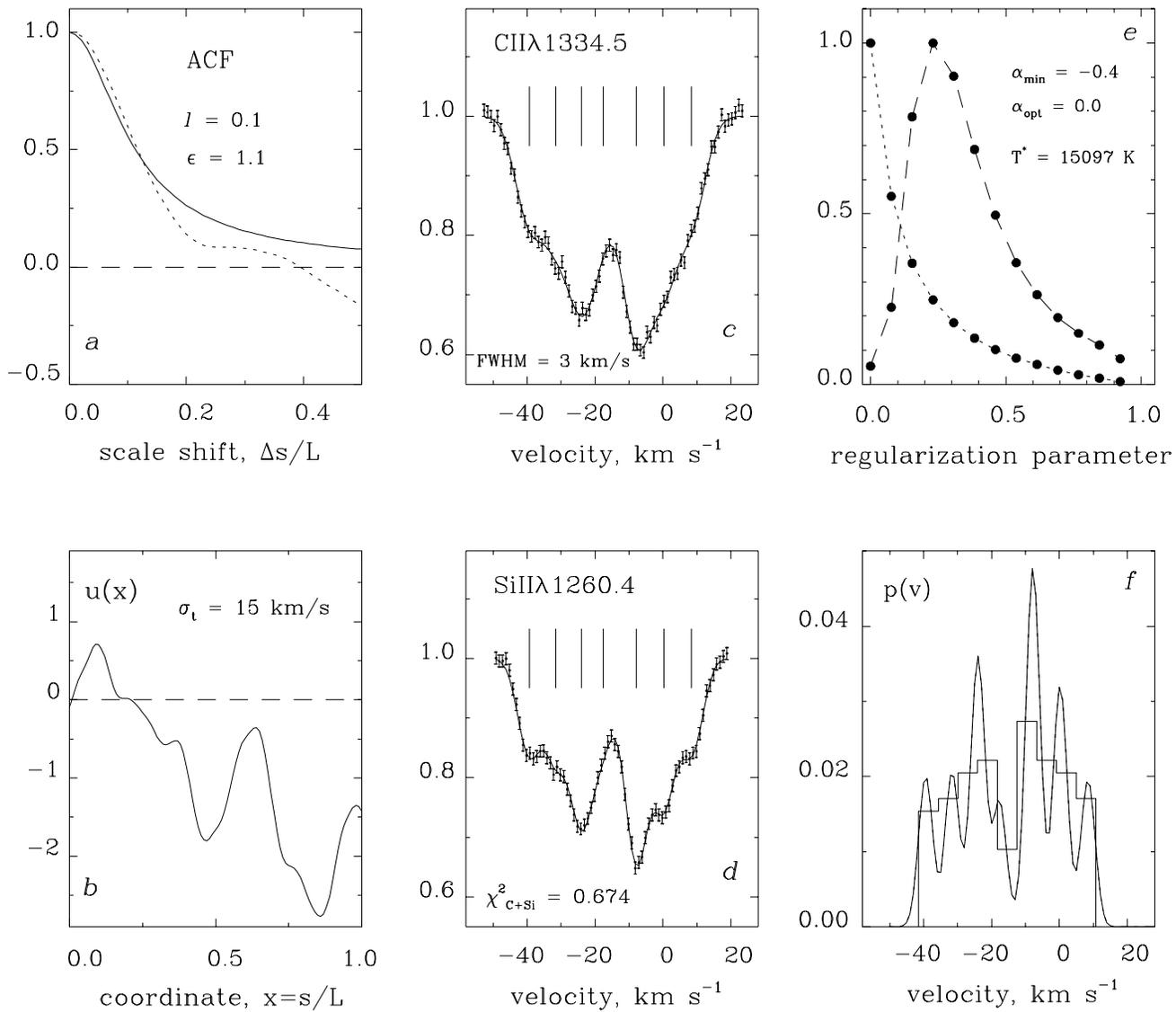
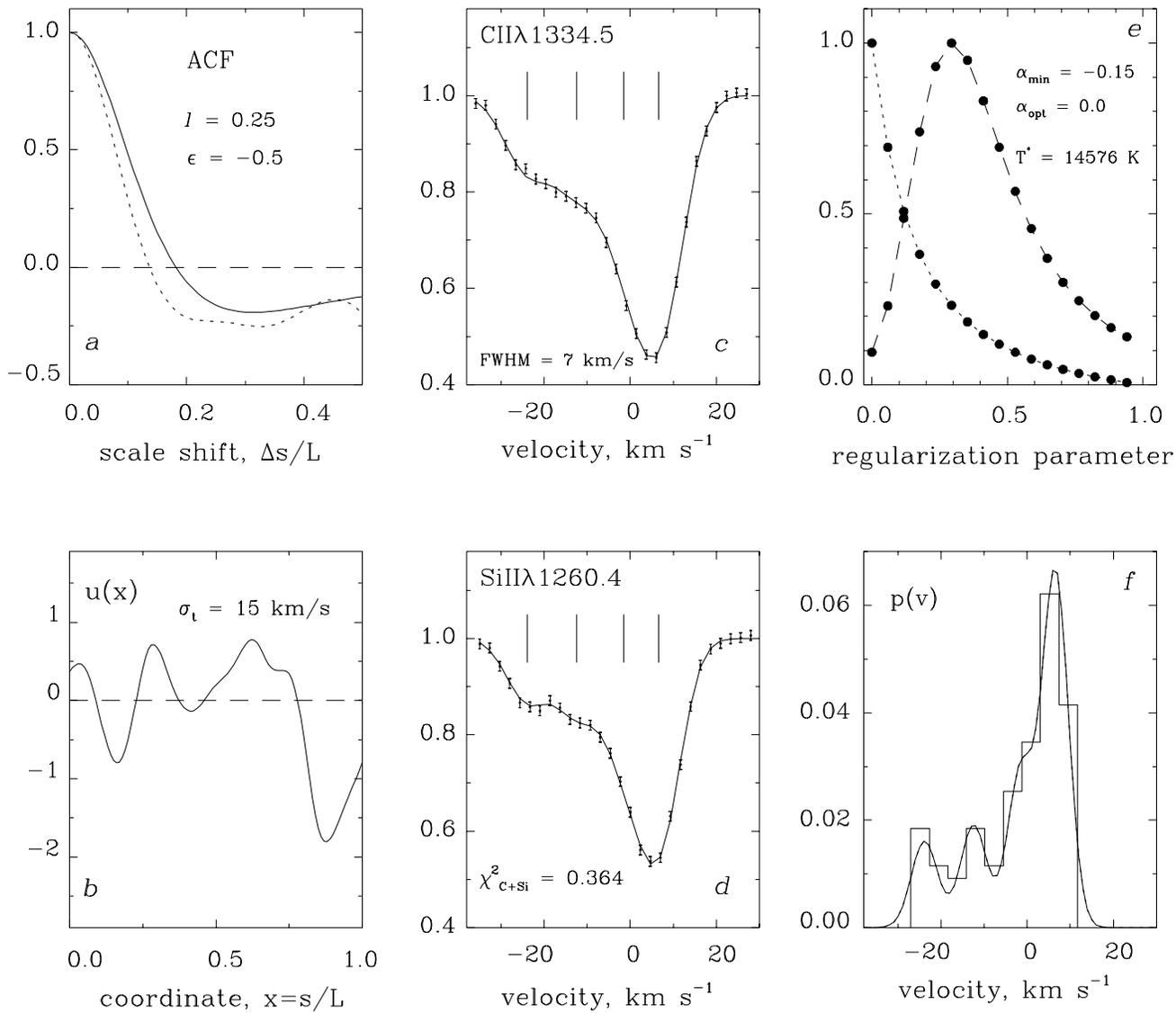
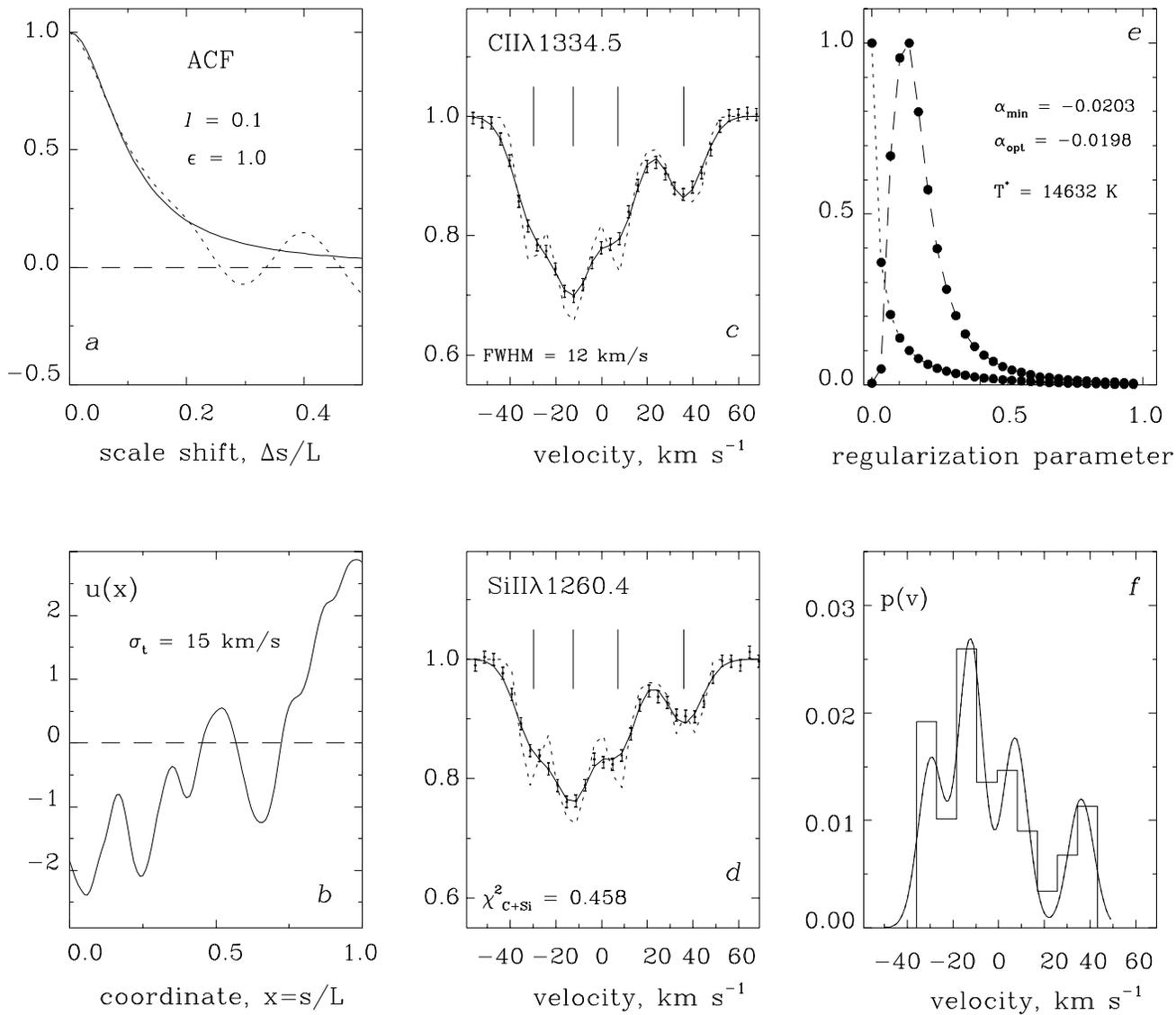
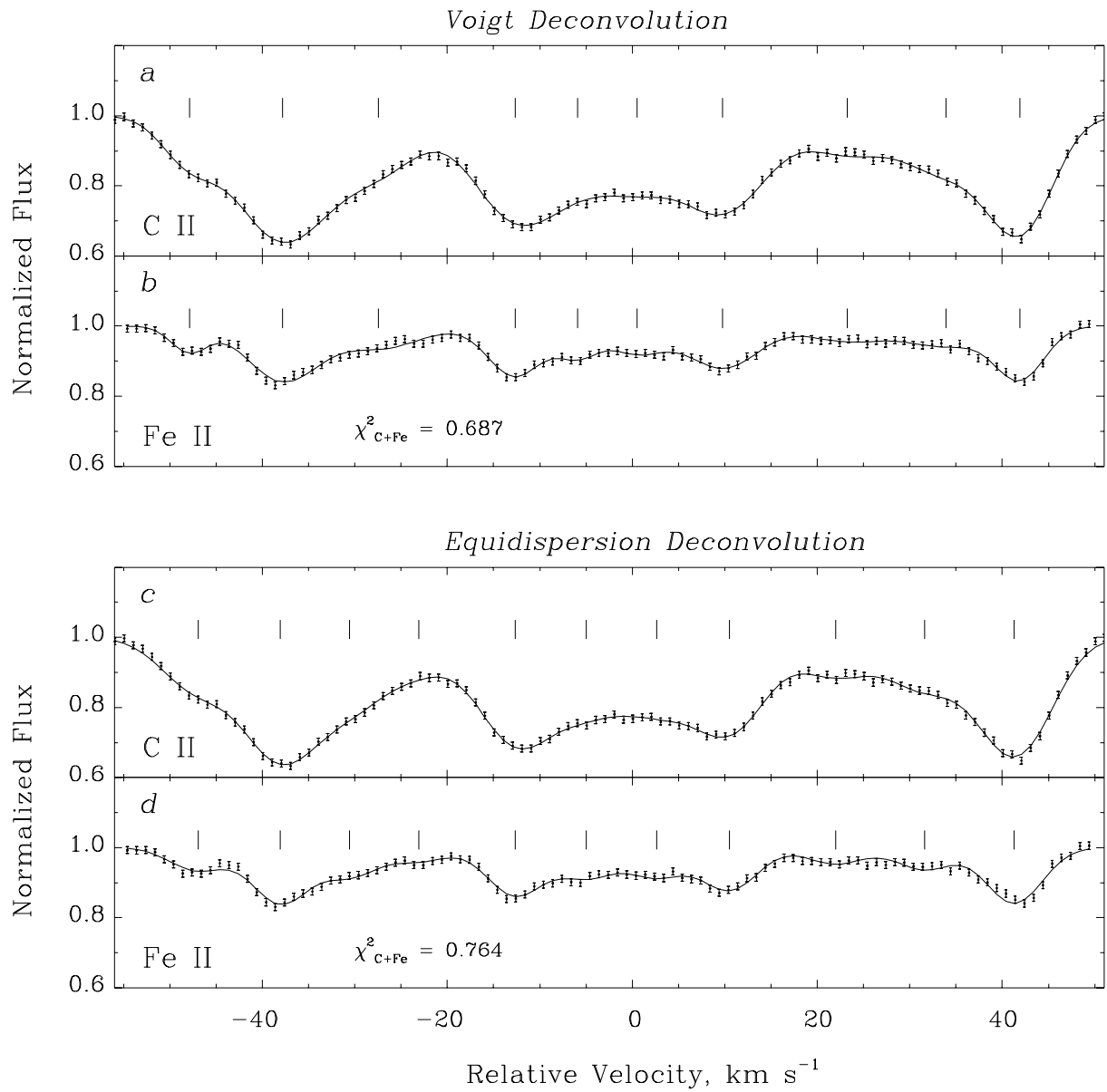
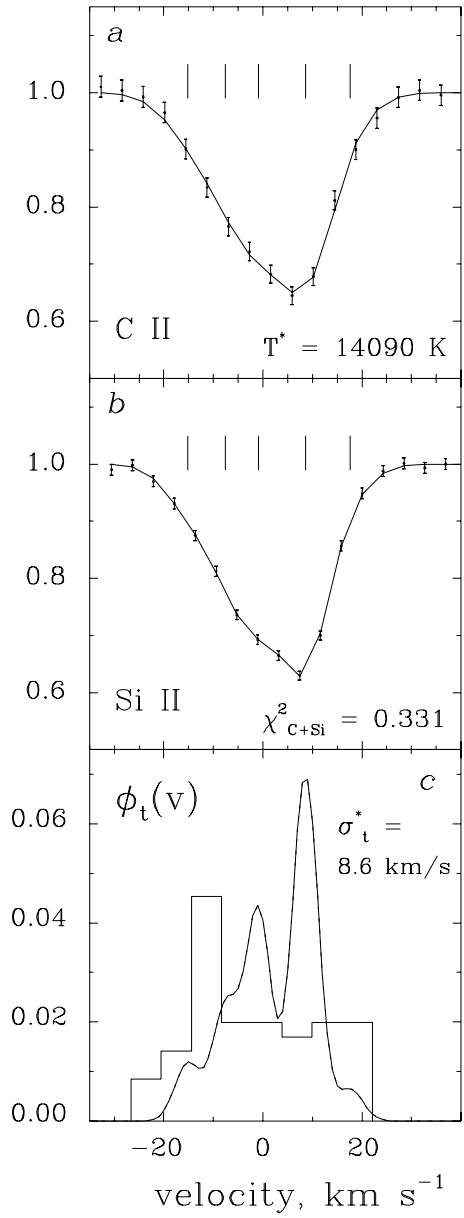
\end{document}